\documentstyle[sprocl,psfig]{article}

\makeatletter
\def\citenum#1{{\def\@cite##1##2{##1}\cite{#1}}}            
\makeatother

\def\be{\begin{equation}}
\def\ee{\end{equation}}
\def\bea{\begin{eqnarray}}
\def\eea{\end{eqnarray}}

\def\qb{\bar{q}}
\def\semi{;\ }
\newskip\humongous \humongous=0pt plus 1000pt minus 1000pt
\def\caja{\mathsurround=0pt}
\def\eqalign#1{\,\vcenter{\openup1\jot \caja
        \ialign{\strut \hfil$\displaystyle{##}$&$
        \displaystyle{{}##}$\hfil\crcr#1\crcr}}\,}
\newif\ifdtup

\newcounter{eqnumber}
\renewcommand{\theeqnumber}{\arabic{eqnumber}}
\def\equn{
\refstepcounter{eqnumber}
\eqno({\rm \theeqnumber})
}
\def\eqn#1{eq.~(\ref{#1})}

\def\fig#1{fig.~{\ref{fig:#1}}}

\def\pol{\varepsilon}

\def\Tr{{\rm Tr}}

\newbox\charbox
\newbox\slabox
\def\s#1{{      
        \setbox\charbox=\hbox{$#1$}
        \setbox\slabox=\hbox{$/$}
        \dimen\charbox=\ht\slabox
        \advance\dimen\charbox by -\dp\slabox
        \advance\dimen\charbox by -\ht\charbox
        \advance\dimen\charbox by \dp\charbox
        \divide\dimen\charbox by 2
        \raise-\dimen\charbox\hbox to \wd\charbox{\hss/\hss}
        \llap{$#1$}
}}
\def\coeff#1#2{\textstyle{#1\over#2}}
\def\I{{\cal I}}
\def\hf{\textstyle{1\over2}}

\def\e{\epsilon}

\def\si{\sigma}

\def\cg{c_{\Gamma}}

\def\tree{{\rm tree}}
\def\oneloop{{\rm 1\! -\! loop}}
\def\twoloop{{\rm 2\! -\! loop}}

\def\qb{\bar q}

\def\jb{{\bar\jmath}}

\def\dlips{d^D{\rm LIPS}}

\def\si{\sigma}

\def\dc{d_{\rm cut}}
\def\yc{y_{\rm cut}}
\def\Split{\mathop{\rm Split}\nolimits}

\def\spa#1.#2{\left\langle#1\,#2\right\rangle}
\def\spb#1.#2{\left[#1\,#2\right]}
\def\lor#1.#2{\left(#1\,#2\right)}
\def\sand#1.#2.#3{%
  \left\langle\smash{#1}{\vphantom1}\right|{#2}%
  \left|\smash{#3}{\vphantom1}\right\rangle}
\def\sandp#1.#2.#3{%
  \left\langle\smash{#1}{\vphantom1}^{-}\right|{#2}%
  \left|\smash{#3}{\vphantom1}^{+}\right\rangle}
\def\sandpp#1.#2.#3{%
  \left\langle\smash{#1}{\vphantom1}^{+}\right|{#2}%
  \left|\smash{#3}{\vphantom1}^{+}\right\rangle}
\def\sandmm#1.#2.#3{%
  \left\langle\smash{#1}{\vphantom1}^{-}\right|{#2}%
  \left|\smash{#3}{\vphantom1}^{-}\right\rangle}
\def\sandpm#1.#2.#3{%
  \left\langle\smash{#1}{\vphantom1}^{+}\right|{#2}%
  \left|\smash{#3}{\vphantom1}^{-}\right\rangle}
\def\sandmp#1.#2.#3{%
  \left\langle\smash{#1}{\vphantom1}^{-}\right|{#2}%
  \left|\smash{#3}{\vphantom1}^{+}\right\rangle}
\begin{document}
\noindent hep-ph/9802264 \hfill {SLAC--PUB--7749}
\par\noindent February 1998  \hfill {CERN--TH/98--39}
\rightline{UCLA/98/TEP/2}
\title{MULTI-PARTON LOOP AMPLITUDES AND NEXT-TO-LEADING ORDER JET 
CROSS-SECTIONS\,\footnote{
Talk presented by L.D. at the International Symposium on QCD Corrections 
and New Physics, Hiroshima, Japan, October 27--29, 1997.
Research supported in part by the US Department of Energy
under grants DE-FG03-91ER40662 (Z.B.) and DE-AC03-76SF00515 (L.D.),
the Laboratory of the {\it Direction des Sciences de la Mati\`ere\/}
of the {\it Commissariat \`a l'Energie Atomique\/} of France (D.A.K.),
the Swiss National Science Foundation (A.S.),
and NATO Collaborative Research Grant CRG--921322 (L.D. and D.A.K.).}}
\author{ZVI BERN}
\address{Department of Physics, University of California, Los Angeles\\
Los Angeles, CA 90024, USA\\E-mail: bern@physics.ucla.edu}
\author{LANCE DIXON}
\address{Stanford Linear Accelerator Center, Stanford University\\ 
Stanford, CA 94309, USA\\E-mail: lance@slac.stanford.edu}
\author{DAVID A. KOSOWER}
\address{Service de Physique Th\'eorique, Centre d'Etudes de Saclay\\
F-91191 Gif-sur-Yvette cedex, France\\E-mail: kosower@spht.saclay.cea.fr}
\author{ADRIAN SIGNER}
\address{Theory Division, CERN, CH-1211 Geneva 23, Switzerland
\\E-mail: Adrian.Signer@cern.ch}
\maketitle
\abstracts{
We review recent developments in the calculation of QCD loop amplitudes
with several external legs, and their application to next-to-leading
order jet production cross-sections.
When a number of calculational tools are combined together 
--- helicity, color and supersymmetry decompositions, 
plus unitarity and factorization properties --- 
it becomes possible to compute multi-parton one-loop QCD 
amplitudes without ever evaluating analytically standard one-loop
Feynman diagrams.   
One-loop helicity amplitudes are now available for  
processes with five external partons ($ggggg$, $q\qb ggg$ and 
$q\qb q'\qb' g$), and for an intermediate vector boson 
$V\equiv\gamma^*,Z,W$ plus four external partons 
($Vq\qb gg$ and $Vq\qb q'\qb'$).  Using these amplitudes, numerical
programs have been constructed for the next-to-leading order corrections 
to the processes $p\bar{p} \to 3$ jets (ignoring
quark contributions so far) and $e^+e^- \to 4$ jets.
}
%


\section{Introduction}

The title of this symposium, ``QCD Corrections and New Physics'',
provides the motivation for this talk:  Many potential signatures
of new physics require a quantitative understanding of Standard Model
backgrounds, which often can only be adequately provided after 
QCD corrections to the lowest order processes have been computed.
The more partons there are in the lowest order process, the more
important the QCD corrections become.  On the other hand, even the first 
corrections, those that are next-to-leading order (NLO) in $\alpha_s$,
have until recently only been available for processes with 
relatively few external partons, such as single-jet and dijet production
at hadron colliders,\cite{NLOTwoJets} and three-jet production in 
$e^+e^-$ annihilation.\cite{eeThreeJets}  

A major reason for this situation has been the dearth of one-loop
QCD amplitudes (virtual corrections) for processes with five or more
external legs.  Such amplitudes are in principle straightforward to
compute from Feynman rules; however, the large number of kinematic
variables means that a brute-force approach can result in extremely large
analytical expressions.  Hence until recently one-loop amplitudes formed
an `analytical bottleneck' to the computation of more Standard Model processes
at next-to-leading order.  This talk will review techniques for
systematically disentangling these amplitudes into components, 
{\it primitive amplitudes}, which are sufficiently simple that they can 
be reconstructed essentially entirely from their analytic properties,
i.e. without resorting to an analytic evaluation of one-loop Feynman
diagrams at all.   

Some of the techniques discussed here have been used to compute
the one-loop five-parton amplitudes, $ggggg$,~\cite{FiveGluon} 
$q\qb q'\qb' g$,~\cite{Kunsztqqqqg} and $q\qb ggg$,~\cite{qqggg}
which enter into NLO studies of the three-jet 
production rate and the substructure of jets at hadron 
colliders.\cite{TThreeJet,KGThreeJet}
More recently, the full set of techniques has led to the one-loop 
helicity amplitudes for a vector boson $V$ (which may be 
a virtual photon, $Z$ or $W$) plus four partons:  
$Vq\qb q'\qb'$~\cite{Zqqqq} and $Vq\qb gg$.~\cite{ZqqggConf,Zqqgg}
The helicity-summed versions of the one-loop corrections to 
$\gamma^*q\qb q'\qb'$ and $\gamma^*q\qb gg$ were calculated contemporaneously
by conventional techniques,\cite{GloverMiller,CGMqggq} although the 
analytic formulae were too large to present.  After converting to a common 
regularization scheme, the two sets of results agree 
numerically.\cite{PrivateNigel}  

The $Vq\qb q'\qb'$ and $Vq\qb gg$ amplitudes have been used to compute the
NLO corrections to $e^+e^-$ annihilation into four
jets.\cite{FourJetsLC,FourJets,NTFourJets} The QCD prediction for the
four-jet rate is almost doubled in going from leading order to NLO,
bringing it into much better agreement with experiment, for several
different jet algorithms.  Besides the total four-jet rate, jet angular
distributions and any other infrared-safe event-shape variable whose
perturbative expansion begins at order $\alpha_s^2$ can now be computed at
NLO.\cite{NTFourJets,SignerMoriond,NTColorFactors} The latter results can
be used to refine experimental measurements of the QCD color factors
$C_F$, $C_A$ and $N_f$.  The same amplitudes could also be applied to NLO
processes that are related by crossing symmetry and coupling constant
conversions, such as the production of a $W$ or $Z$ plus two jets at
hadron colliders, or the deep inelastic production of three jets, but this
has not yet been carried out.

Many aspects of the analytical techniques covered here have been previously 
reviewed elsewhere.\cite{AnnualReview,TASIReview,OtherReviews,Zqqgg}
In particular, the recent calculation of $e^+e^- \to V \to q\qb gg$
required a thorough understanding of the spurious as well as physical
singularities that can occur in six-point amplitudes at one loop; such
details are beyond the scope of this review.


\section{Primitive Amplitudes and Analytic Properties}

\subsection{Traditional Approaches}

In the traditional approach to computing a gauge theory amplitude, one 
begins by drawing all Feynman diagrams.  However, individual diagrams
are not gauge invariant; only certain sums of diagrams are.
Thus each diagram contains unphysical, gauge-variant pieces, which
one would like to avoid calculating at all, since such pieces must cancel
eventually.  Furthermore, the standard Feynman rules for the
gauge boson self-interactions induce a complicated tangle of color 
and Lorentz indices in each diagram.  The three-gluon vertex is
$$
V^{abc}_{\mu\nu\rho}(k,p,q) = f^{abc} \Bigl(
   \eta_{\nu\rho}(p-q)_\mu
+ \eta_{\rho\mu} (q-k)_\nu + \eta_{\mu\nu} (k-p)_\rho \Bigr) \,,
\equn\label{FeynmanVertex}
$$
where $f^{abc}$ are the $SU(3)$ structure constants, $k$, $p$ and $q$
the momenta emerging from the vertex, and $\eta_{\mu\nu}$ the 
Minkowski metric.  Each of the three terms in \eqn{FeynmanVertex}
leads to a different routing of the Lorentz indices, and the
composition of several $f^{abc}$s similarly leads to different routings of
the color indices.

In a loop amplitude, factors of loop momenta coming from
the vertices lead to further complications when one attempts to reduce
the various loop integrals to a basic set of integral functions.
The pentagon integral appearing in the one-loop five-gluon amplitude
includes many terms with five powers of loop momenta in the numerator, 
arising from five vertices of the form~(\ref{FeynmanVertex}).
Each such term can amount to a page or more of algebra after reduction
of the integral.

The complications of a diagrammatic approach are in marked contrast to the
simplicity of the final results.  For example, in the mid-1980s it was found
that all QCD tree amplitudes with four or five external partons (and many
with six or more external partons) could be written as permutation sums 
of simple, single-term expressions 
(see \eqn{FiveGluonTree}).\cite{ManganoReview}
Similarly, the one-loop four- and five-gluon amplitudes can be
represented remarkably compactly 
(see \eqn{FiveGluonLoop}).\cite{Long,KunsztFourPoint,FiveGluon}


\subsection{Outline of Analytic Approach}

This contrast suggests that there should be a more direct and
manifestly gauge-invariant way to calculate loop amplitudes than via
Feynman diagrams.  Here we shall advocate exploiting the two principal
analytic properties of loop amplitudes, unitarity and factorization on
particle poles, in a `perturbative bootstrap' approach.  In the old
$S$-matrix bootstrap program of the 1960s,\cite{Bootstrap} one
attempted to construct $S$-matrices for the strong interactions using
the constraints of analyticity.  However, in the absence of a
perturbative expansion, these nonlinear, coupled constraints proved
too difficult to solve exactly, without imposing artificial
constraints of some kind, for example on the number of states in the
physical spectrum.  On the other hand, in the context of perturbation
theory, the constraints of $S$-matrix analyticity become {\it
recursive} in the coupling constant $g$, and therefore tractable.  

\begin{figure}[t]
\psfig{figure=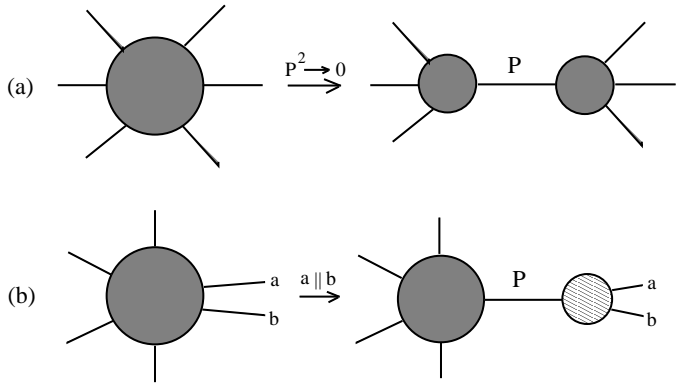,height=2in}
\caption{
(a) Factorization of a tree amplitude on a particle pole, into 
the product of an $(m+1)$-point tree amplitude and an $(n-m+1)$-point 
tree amplitude.
(b) The case $m=n-2$ (or $m=2$) corresponds to two external legs,
$a$ and $b$, becoming collinear.  Gray blobs represent tree amplitudes,
while the hatched blob represents a splitting amplitude.
}  
\label{fig:FactorizationFigure}
\end{figure}

For example, an $n$-point tree amplitude is said to factorize on an 
intermediate particle pole when the sum of $m$ external momenta
satisfies the on-shell condition for an intermediate state.  
The factorization condition for the massless case is $P^2 \to 0$, 
where $P \equiv k_{i_1}+k_{i_1+1}+\cdots+k_{i_2}$.
In this limit, depicted in \fig{FactorizationFigure}(a), the
$n$-point amplitude becomes
$$
\eqalign{
&A^\tree_n(1,2,\ldots,n) \cr
&\mathop{\longrightarrow}^{P^2 \to 0}\  
  A^\tree_{m+1}(i_1,\ldots,i_2,(-P)^{-\lambda})\ {1\over P^2}\ 
  A^\tree_{n-m+1}(P^\lambda,i_2+1,\ldots,i_1-1) \,, \cr
}\equn\label{PoleEquation}
$$
where we take all momenta to be outgoing in each amplitude, and there
is an implicit sum over intermediate polarization states $\lambda$.
(We have suppressed the color indices here, as is appropriate for the
color-ordered primitive amplitudes defined below.)
The special case $m=n-2$ (or $m=2$), shown in \fig{FactorizationFigure}(b), 
occurs when two of the external momenta become collinear, say
$k_a \parallel k_b$:
$$
A^\tree_n(\ldots,a,b,\ldots)\ \mathop{\longrightarrow}^{a \parallel b}\
 \Split^\tree_{-\lambda}(a^{\lambda_a},b^{\lambda_b})\,
      A^\tree_{n-1}(\ldots,P^\lambda,\ldots) \,.
\equn\label{CollinearEquation}
$$
The collinear singularity behaves like the 
square-root of a pole, which has been absorbed into a universal 
{\it splitting amplitude} that replaces the second amplitude  
on the right-hand side of \eqn{PoleEquation}. 
 
The universal structure of these tree-level limits is easy to
understand heuristically:  The spacetime picture of the scattering 
process reduces to two independent scatterings connected by a single
intermediate state.  It can also be derived from a string-theoretic
representation of the tree amplitudes.\cite{ManganoReview}
In any case, the `boundary
values' (factorization limits) of the $n$-point amplitude contain only
lower-point amplitudes or universal functions.   Thus the general 
$n$-point tree amplitude could be constructed recursively
in $n$, simply by demanding consistency with all factorization limits,
provided only that the solution to the boundary-value problem
is unique.   (We will discuss the uniqueness issue later.)

In fact, tree amplitudes can be constructed rather efficiently, and
without any uniqueness issues, using recursive techniques based
on Feynman diagrams.\cite{RecursiveBG,Recursive}
However, for loop amplitudes this is not generally the case,
and so the factorization properties of 
loop amplitudes\,\cite{SusyFour,BCFactorization} become quite useful.
These properties are straightforward generalizations of the corresponding
tree properties.  For example, the generic form of the collinear limit of 
a one-loop amplitude is  
$$
\eqalign{
A^\oneloop_n(\ldots,a,b,\ldots)
\ \mathop{\longrightarrow}^{a \parallel b}\
 &\Split^\tree_{-\lambda}(a^{\lambda_a},b^{\lambda_b})\,
      A^\oneloop_{n-1}(\ldots,P^\lambda,\ldots) \cr
&+ \Split^\oneloop_{-\lambda}(a^{\lambda_a},b^{\lambda_b})\,
      A^\tree_{n-1}(\ldots,P^\lambda,\ldots) \,, \cr
}\equn\label{LoopCollinearEquation}
$$
where $\Split^\oneloop$ is a new universal function.

\begin{figure}[t]
\psfig{figure=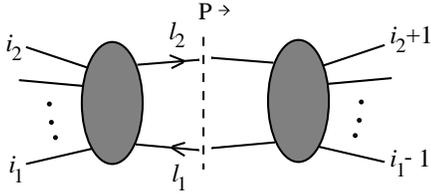,height=1in}
\caption{
A cut for a one-loop amplitude, with momentum $P$ flowing across the cut. 
The blobs on either side of the cut represent tree amplitudes.
}  
\label{fig:CutAmplitudeFigure}
\end{figure}

Even more useful for loop amplitudes are the constraints imposed by 
unitarity.  After expanding the $S$-matrix unitarity relation 
$S^\dagger S=1$ in terms of $g$, one finds that the cut or absorptive
part of a one-loop $n$-point amplitude is given by the two-body 
phase-space integral of a product of two tree amplitudes,
$$
\eqalign{
A^\oneloop_n (1, 2,\ldots,n)
       \Bigr|_{ P^2 \ \rm cut} \!\!
&= i \int \dlips(-\ell_1,\ell_2)\ 
A^\tree_{m+2}((-\ell_1)^{-\lambda_1},i_1,\ldots,i_2,\ell_2^{\lambda_2}) \cr
&\hskip1.0cm \times 
A^\tree_{n-m+2}((-\ell_2)^{-\lambda_2},i_2+1,\ldots,i_1-1,\ell_1^{\lambda_1})
\,, \cr}
\equn\label{CutEquation}
$$
as illustrated in \fig{CutAmplitudeFigure}.
Here the cut channel has momentum squared $P^2$,
the integration is over $D$-dimensional Lorentz-invariant
phase space with with momenta $-\ell_1$ and $\ell_2$ for the
intermediate states, and there is again an implicit sum over their
polarizations $\lambda_1,\lambda_2$.  For all the one-loop multi-parton 
amplitudes that one might presently contemplate calculating, the tree 
amplitudes required to evaluate each of the cuts are readily available
in a compact form.

\subsection{Decomposition into Primitive Amplitudes}

Although the above factorization and unitarity constraints are powerful,
they are somewhat cumbersome to implement on full multi-parton gauge 
theory amplitudes.  It is much simpler to first identify a simpler
set of gauge-invariant building blocks, 
called {\it primitive amplitudes},\cite{qqggg}
from which the full amplitudes can be built, but which individually
have simpler analytic properties.  The two handles that one can use to 
pull apart QCD amplitudes are the {\it helicity} and {\it color}
quantum numbers of the external quarks and gluons.

In massless QCD, quark helicity is conserved, and decomposing an amplitude
with respect to it just amounts to inserting a helicity projector
$\hf(1\pm\gamma_5)$ next to the external spinor.  A definite gluon helicity
is simplest to implement via the spinor helicity 
formalism,\cite{SpinorHelicity} which represents gluon polarization
vectors in terms of Weyl spinors $|\,k^\pm \rangle$,
$$
\pol^{+}_\mu (k;q) = 
     {\sandmm{q}.{\gamma_\mu}.k
      \over \sqrt2 \spa{q}.k}\, ,\hskip 1cm
\pol^{-}_\mu (k;q) = 
     {\sandpp{q}.{\gamma_\mu}.k
      \over \sqrt{2} \spb{k}.q} \,.
\equn\label{HelicityDef}
$$  
Here $k$ is the gluon momentum, $q$ is an arbitrary null
`reference momentum' which drops out of final gauge-invariant
amplitudes, and the spinor inner-products are
$$
\spa{i}.j \equiv  \langle k_i^{-} \vert k_j^{+} \rangle\,, 
\hskip 1.0 cm 
\spb{i}.j \equiv \langle k_i^{+} \vert k_j^{-} \rangle \,, 
\hskip 1.0 cm
\spa{i}.j \spb{j}.i = s_{ij} = (k_i+k_j)^2 \,. 
\equn\label{AngleDef}
$$ 
 
The color structure of $SU(N_c)$ gauge theory amplitudes can be sorted into
{\it color-ordered} components --- sums of contributions from
Feynman graphs where the external states maintain a definite cyclic
ordering with respect to each other, and where the {\it color-ordered}
vertices have had the color factors removed from them.\cite{AnnualReview}
The color factors that multiply these components, in the color
decomposition of the full amplitude, are ordered traces of generators 
$T^a$ for the fundamental representation of $SU(N_c)$,
$\Tr(1\ldots n) \equiv \Tr(T^{a_1}\ldots T^{a_n})$,
or strings of the form $(T^{a_3}\cdots T^{a_n})_{i_2}^{~\jb_1}$
if external quarks are also present.
For example, the color decompositions of the tree-level and one-loop 
five-gluon amplitudes are
$$
{\cal A}^\tree_5 = g^3 \sum_{\sigma\in S_5/Z_5} 
\Tr(\si(1)\ldots\si(5)) A^\tree_5(\si(1),\ldots,\si(5))  
\equn\label{TreeAmplitude}
$$
and 
$$
\eqalign{
{\cal A}_{5}^\oneloop = &\ g^5 \Biggl[ \,
\sum_{\sigma \in S_5/Z_5}
N_c \, \Tr(\si(1)\ldots\si(5))\  
    A_{5;1} (\si(1),\ldots,\si(5)) \cr
& \hskip-16mm + \hskip-4mm
  \sum_{\sigma \in S_5/(S_2\times S_3)} \hskip-6mm
  \Tr(\si(1)\si(2)) \, \Tr(\si(3)\si(4)\si(5))\ 
A_{5;3} (\si(1),\si(2);\si(3),\si(4),\si(5)) \Biggr] \,, \cr}
\equn\label{LoopColor}
$$
where the permutation sums run over all inequivalent traces. 
$A^\tree_5(1,2,3,4,5)$ and $A_{5;1}(1,2,3,4,5)$ are color-ordered 
amplitudes, so that when their external states are chosen to be of 
definite helicity, e.g. 
$A_{5;1}(1^{\lambda_1},2^{\lambda_2},3^{\lambda_3},4^{\lambda_4},
5^{\lambda_5})$, they qualify as primitive amplitudes.
The same is not quite true of the coefficients $A_{5;3}$ of
the double-trace terms in \eqn{LoopColor}; however,
they can be expressed\,\cite{SusyFour} as sums of permutations of the
$A_{5;1}$s.

The virtue of these decompositions becomes apparent when the explicit
values of $A_5^\tree$ are recorded:
$$
\eqalign{
A_5^\tree(1^\pm,2^+,3^+,4^+,5^+)\ &=\ 0, \cr
A_5^\tree(1^-,2^-,3^+,4^+,5^+)\ &=\ 
i\, {\spa1.2^4\over\spa1.2\spa2.3\spa3.4\spa4.5\spa5.1}\ , \cr
A_5^\tree(1^-,2^+,3^-,4^+,5^+)\ &=\ 
i\, {\spa1.3^4\over\spa1.2\spa2.3\spa3.4\spa4.5\spa5.1}\ ; \cr
}\equn\label{FiveGluonTree}
$$
the remaining helicity configurations are related by parity and charge
conjugation.  

The zeros in \eqn{FiveGluonTree} are a general consequence
of supersymmetric Ward identities (SWI).\cite{OldSWI}
At tree level, SWI can be applied directly to a non-super\-symmetric theory 
like QCD, because no fermions can contribute to a tree-level 
$n$-gluon amplitude; hence the `missing' fermions might as well be 
considered gluinos instead of quarks.\cite{NewSWI}

The simple analytic structure of the nonzero terms 
is due to the color and helicity decompositions.  
Color-ordering implies that factorization
singularities can only appear in color-adjacent channels.
Thus only a subset of the kinematic variables 
--- here, $\{ s_{12},s_{23},s_{34},s_{45},s_{51} \}$ ---
are `important' in the sense that they contain all poles (and in the loop
case, all cuts) in the primitive amplitude.
The denominator factors in \eqn{FiveGluonTree},
$\spa{i}.{(i+1)} = (\hbox{phase}) \times \sqrt{s_{i,i+1}}$, 
properly capture the square-root behavior of collinear singularities, 
including a phase dependence as $k_i$ and $k_{i+1}$ are rotated around their 
common collinear axis; both of these factors are related to an 
angular-momentum mismatch of 1 unit in the collinear limit.  
(In other cases, $\spb{i}.{(i+1)}$ denominator factors can appear, when the
mismatch has the opposite sign.) 

One-loop primitive amplitudes benefit in precisely the same way as tree
amplitudes from color and helicity decompositions.  In addition, it is
possible to decompose a one-loop amplitude according to the spins of the
particles going around the loop, in a `supersymmetric' way.  For example,
for a QCD amplitude with all external gluons, the internal gluon loop
contribution $g$ (and fermion loop contribution $f$) can be rewritten as a
supersymmetric contribution plus a complex scalar loop $s$,
$$
\eqalign{
  g\ &=\ (g+4f+3s)\ -\ 4(f+s)\ +\ s\ =\ A^{N=4}\ -\ 4\,A^{N=1}
  \ +\ A^{\rm scalar}, \cr
  f\ &=\ (f+s)\ -\ s\ =\ A^{N=1}\ -\ A^{\rm scalar}, \cr
}\equn\label{SusyDecomp}
$$
where $A^{N=4}$ represents the contribution of the $N=4$ super Yang-Mills
multiplet, and $A^{N=1}$ an $N=1$ chiral matter supermultiplet. 
(Related rearrangements are possible for amplitudes with external 
quarks.\cite{qqggg,Zqqgg})
The two supersymmetric contributions, $A^{N=4}$ and $A^{N=1}$,
automatically obey SWI, and have other simplifications arising from 
loop-momentum cancellations.
The scalar contribution is generally the most algebraically complicated 
of the three, but it is also simpler in some aspects.

We illustrate this decomposition for $A_{5;1}(1^-,2^-,3^+,4^+,5^+)$,
which is one of the two one-loop primitive amplitudes required to calculate
the `gluonic' NLO corrections to $p\bar{p} \to 3$~jets. 
Its components according to~(\ref{SusyDecomp}) are\,\cite{FiveGluon} 
$$
\eqalign{
  A^{N=4} &= \cg A^\tree_5 \sum_{j=1}^5 \Biggl[ 
     -{1\over\e^2} \left( {\mu^2\over -s_{j,j+1}}\right)^\e \!\!
     + \ln\left({-s_{j,j+1}\over -s_{j+1,j+2}}\right)
       \ln\left({-s_{j+2,j-2}\over -s_{j-2,j-1}}\right)
     + {\pi^2\over6} \Biggr] \cr
  A^{N=1} &= \cg A^\tree_5 \Biggl[ {5\over2\e} 
    +{1\over2}\left[\ln\left({\mu^2\over -s_{23}}\right)
                +\ln\left({\mu^2\over -s_{51}}\right)\right] + 2 \Biggr] \cr
&\quad + {i\cg\over2}
   {{\spa1.2}^2 \left(\spa2.3\spb3.4\spa4.1+\spa2.4\spb4.5\spa5.1\right)\over
    \spa2.3\spa3.4\spa4.5\spa5.1}
     {\ln\left( {-s_{23}\over -s_{51}}\right)\over s_{51}-s_{23}} \cr
}
$$
$$
\eqalign{
  A^{\rm scalar} &= {1\over3} A^{N=1} 
    + {i\cg\over3} \Biggl[ {2\over3} {A^\tree_5\over i} \cr
&\hskip-11mm 
   - { \spb3.4\spa4.1\spa2.4\spb4.5
       \left(\spa2.3\spb3.4\spa4.1+\spa2.4\spb4.5\spa5.1\right)
          \over\spa3.4\spa4.5 } 
     { \ln\left({-s_{23}\over -s_{51}}\right)
         -{1\over2}\left({s_{23}\over s_{51}}-{s_{51}\over s_{23}}\right)
               \over (s_{51}-s_{23})^3 } \cr
& \hskip-11mm
   - {\spa3.5{\spb3.5}^3\over\spb1.2\spb2.3\spa3.4\spa4.5\spb5.1}
   + {\spa1.2{\spb3.5}^2\over\spb2.3\spa3.4\spa4.5\spb5.1}
   + {1\over2}{\spa1.2\spb3.4\spa4.1\spa2.4\spb4.5\over
                  s_{23}\spa3.4\spa4.5 s_{51}} \Biggr]\ , \cr
}\equn\label{FiveGluonLoop}
$$
where $c_\Gamma = \Gamma(1+\e)\Gamma^2(1-\e)/((4\pi)^{2-\e}\Gamma(1-2\e))$. 
Since the three components have quite different analytic
structure, the rearrangement~(\ref{SusyDecomp}) is a natural one.  
As expected, the $A^{N=4}$ is the simplest component, followed by  
$A^{N=1}$.  While $A^{\rm scalar}$ is the most complicated piece, 
it has (for example) no $s_{34}$ cut, while the full gluon loop does.

Although we do not have space to demonstrate it explicitly here,
evaluation of the unitarity cuts, \eqn{CutEquation}, for an amplitude 
such as \eqn{FiveGluonLoop} is quite simple.  The principle reasons are:
\par\noindent
(1) The tree amplitudes are fully `processed', in terms of gauge
cancellations, etc., before they are fed into the cut.
\par\noindent
(2) SWI generally apply to the tree amplitudes, simplifying their form,
even if they do not directly apply to the one-loop amplitudes.
\par\noindent
(3) On-shell conditions for the intermediate legs, 
$\ell_1^2 = \ell_2^2 = 0$, can be used repeatedly to simplify the 
evaluation.

The only catch to a unitarity-based approach is that loop amplitudes can 
have rational-function terms, containing no logarithms or dilogarithms,
which are not detectable in any cut (when the cut is evaluated in 
four-dimensions).  In \eqn{FiveGluonLoop},
such terms appear only in the non-super\-symmetric component 
$A^{\rm scalar}$.   (The rational-function terms in the supersymmetric
components can be shown quite generally to be linked to the 
logarithms.\cite{SusyOne})  Fortunately these terms can be determined
using the factorization properties of loop amplitudes, such as
\eqn{LoopCollinearEquation}.  One can construct an ansatz for the 
rational-function terms, and then proceed channel by channel to 
add terms to the ansatz so that it correctly reproduces the desired 
singular behavior in each channel, using factorization information 
provided by lower-point amplitudes.  This procedure could fail
if the factorization properties did not uniquely fix the amplitude.
Although we have no formal uniqueness proof, empirically the 
procedure always works for six-point amplitudes.  
We can verify that it has worked by numerical evaluating conventional 
Feynman diagrams at one or more random kinematic points.
(At the five-point level, there is a special term that provides a 
counter-example to uniqueness,\cite{AnnualReview} 
although in practice its coefficient could also be fixed numerically.) 

The above strategy of calculating primitive amplitudes via their unitarity
cuts, with the rational-function terms fixed via factorization, has been
successfully applied to the one-loop $e^+e^- \to \gamma,Z \to 4$~parton
amplitudes.  Most of the previously calculated one-loop four- and 
five-parton amplitudes reappear in this calculation as the 
boundary-value information for the rational-function terms, 
thus emphasizing the bootstrap nature of the approach.


\section{Phenomenological Applications}

\subsection{$p\bar{p} \to 3$~jets}

To date there have been two phenomenological applications of the one-loop
helicity amplitudes whose calculation was outlined above.  First,
two groups have undertaken to compute NLO corrections to three-jet
production in $p\bar{p}$ collisions.\cite{TThreeJet,KGThreeJet}
Production of three jets is one of the dominant processes at large 
transverse momentum at the Tevatron, after the one-jet inclusive and 
two-jet rates which have already been computed at NLO.\cite{NLOTwoJets}
By measuring the three-jet to two-jet ratio, once the former has been 
computed at NLO, one should be able to extract a relatively precise value 
for $\alpha_s(Q^2)$, at the highest experimentally accessible 
values of $Q^2$.

Although all the matrix elements required for the full
NLO $p\bar{p} \to 3$~jets calculation are available,
the extreme computational demands of the numerical integrals
encountered in the calculation have led both groups to perform 
a `gluonic approximation' to the true result, 
in which the quark distribution functions for the proton are
dropped, as well as final-state quarks.  Thus only the one-loop
five-gluon matrix elements\,\cite{FiveGluon} and tree-level six-gluon
matrix elements\,\cite{TreeSixGluon,RecursiveBG,ManganoReview} have to be 
evaluated.  Although this `approximation' is not expected to be
particularly good quantitatively, it provides some useful qualitative lessons.

\begin{figure}[t]
\psfig{figure=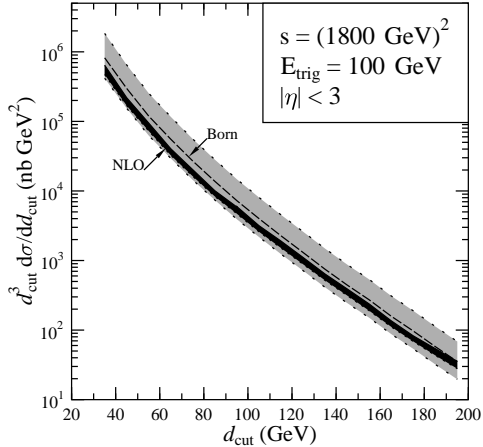,height=3in}
\caption{
Dependence of the gluonic $p\bar{p}\to 3$~jet cross-section
on the jet resolution parameter $\dc$ for the $k_T$ clustering 
algorithm.\protect\cite{TThreeJet}
The gray and black bands result from varying
the renormalization/factorization scale $\mu$ over the range 
$\dc/2<\mu<2\dc$ for the
Born and NLO results, respectively.
}    
\label{fig:TThreeJetFigure}
\end{figure}

Tr\'ocs\'anyi\,\cite{TThreeJet} used an iterative 
$k_T$ clustering algorithm\,\cite{ppkt}
adapted from $e^+e^-$ annihilation, with a variable jet resolution
parameter $\dc$ which controls the overall hardness of the three-jet event.
The $\dc$-dependence of the leading-order (Born) and NLO
results are shown in \fig{TThreeJetFigure}.  For this jet algorithm, the 
Born and NLO cross-section have very similar $\dc$-dependence.
However, the NLO result has (as expected) a much reduced dependence on the
(unphysical) renormalization scale $\mu$, which bodes well for the
accuracy of the full NLO three-jet rate once quark contributions are included. 

Kilgore and Giele\,\cite{KGThreeJet} studied $p\bar{p} \to 3$~jets 
for the $k_T$ algorithm as well as for three algorithms of the cone
type more commonly used at hadron colliders:
\par\noindent
1) a `fixed cone' algorithm, used by UA2,\,\cite{FixedCone}
\par\noindent
2) an `iterative cone' algorithm, used by CDF and D0,\,\cite{IterativeCone} 
\par\noindent
3) the `EKS' algorithm, used in NLO one- and two-jet inclusive 
calculations.\cite{EKSalg}
\par
\begin{figure}[t]
\psfig{figure=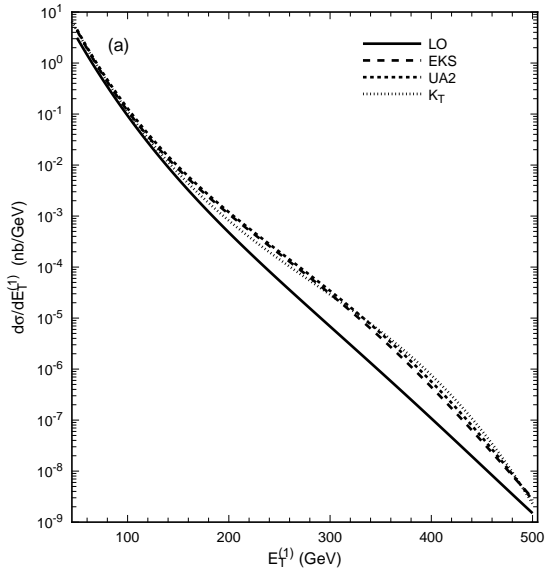,height=3in}
\caption{
The NLO gluonic $p\bar{p}\to 3$~jet cross-section 
as a function of the transverse energy of the leading jet, $E_T^{(1)}$, 
for three different jet algorithms and $\mu = 100$~GeV, 
from ref.~\protect\citenum{KGThreeJet}.   See text for descriptions of the
algorithms.  The Born (LO) result is also shown for comparison.
}  
\label{fig:KGThreeJetFigure}
\end{figure}

\noindent
Their NLO results for all but the iterative cone algorithm are shown in
\fig{KGThreeJetFigure}.  The dependence of the three-jet
cross-section on the transverse energy of the leading jet, $E_T^{(1)}$, 
changes considerably between leading and next-to-leading order;
it also varies with the jet algorithm.  

There is no plot for the iterative cone algorithm --- the one
presently used by CDF and D0 --- because Kilgore and Giele found it to
be infrared unsafe for the three-jet cross-section.  
(See also ref.~\citenum{SeymourIR}.)  They traced this
problem to final states with three hard partons, two of which are
separated by slightly more than the cone size $R$.  If a soft gluon is
added to this configuration somewhere between these two hard partons,
its only effect in the other jet algorithms will be to shift one of
the three jet axes slightly.  But in the iterative cone algorithm this
shift will (in a subsequent step) trigger the merging of two of the
three jets into one, thus altering the three-jet rate.  (In fact, CDF
and D0 had already introduced an additional jet separation cut when
comparing their multi-jet data to leading order predictions, which
effectively removed this problem, at least at NLO.)  
The full $p\bar{p}\to 3$~jet NLO
cross-section still awaits the inclusion of quarks; however, the
infrared dangers of the iterative cone algorithm already constitute a
useful lesson from the gluon-only study.


\subsection{$e^+e^- \to 4$~jets}

A second phenomenological application has been in $e^+e^-$ annihilation.
Here four-jet production, or more generally infrared-safe
event-shape variables whose perturbative expansion begins at 
order $\alpha_s^2$, can now be studied to NLO accuracy. 
Such observables are sensitive to the non-Abelian self-coupling of gluons 
and to the production of hypothetical colored, electrically neutral 
particles such as light gluinos.
Also, at LEP2 energies $e^+e^-\to \gamma,Z \to 4$~jets forms a 
significant background to $W$ pair production when both $W$s decay 
hadronically.   

To date, two independent numerical programs for 
NLO corrections to generic order $\alpha_s^2$ event shapes have been
constructed, MENLO\_PARC\,\cite{MENLOPARC} and DEBRECEN,\cite{NTFourJets}
implementing the one-loop matrix elements for 
$e^+e^-\to \gamma,Z \to$ 4 partons, and the tree-level matrix elements for 
$e^+e^-\to \gamma,Z \to$~5 partons.\cite{BerGie89,FiveJetsBorn}
Although the two programs use different generalizations of the subtraction 
method\,\cite{ERT,KunsztSoperFKSCS} for carrying out integrals over the 
singular five-parton phase-space, their results agree to within
statistical errors for the Monte Carlo integration.  

\begin{figure}[t]
\psfig{figure=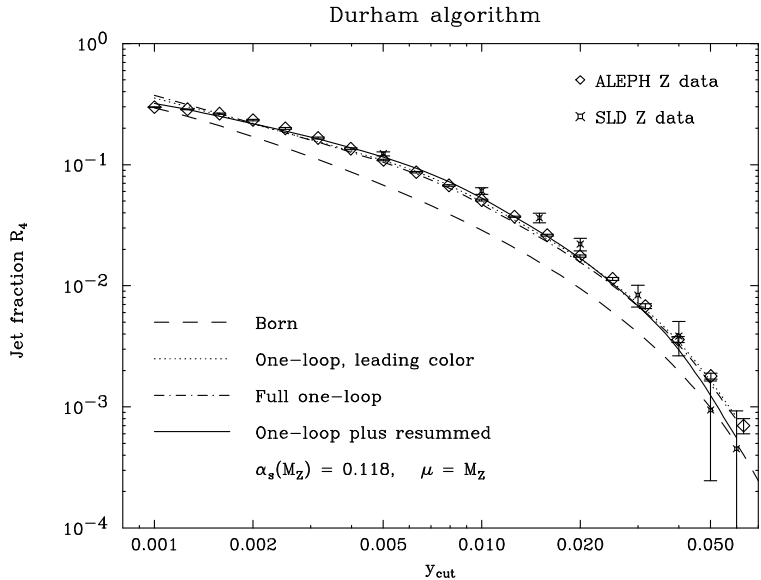,height=3in}
\caption{
The four-jet fraction in $e^+e^-$ annihilation for the Durham jet
algorithm, from ref.~\protect\citenum{FourJets}.   
(A small numerical error in the resummed curve has been corrected here.)
}  
\label{fig:eeFourJetFigure}
\end{figure}

In \fig{eeFourJetFigure}, QCD predictions\,\cite{FourJets} 
for the four-jet fraction $R_4$ using the Durham jet 
algorithm\cite{Durham,DurhamResum} are compared 
with data at the $Z$ pole
from ALEPH\,\cite{ALEPHReview} and SLD,\cite{SLDdata} 
as a function of the jet resolution parameter $\yc$.
It has long been known that
the Born-level QCD prediction for the Durham (and most other $e^+e^-$ jet) 
algorithms, when evaluated at a renormalization scale 
$\mu=\sqrt{s}$, under-predicts the measured four-jet fraction by roughly a
factor of two.  The figure shows that addition of the NLO
correction brings the prediction into much better agreement with the data,
within roughly 10\%.   

For small values of $\yc$, where the four-jet fraction is large,
a kinematic singularity develops in the perturbative expansion,
and the true expansion parameter becomes $\alpha_s L^2$, where 
$L = \ln(1/\yc)$.  The Durham algorithm has the virtue of being 
resummable --- the leading and next-to-leading logarithms at each order
in $\alpha_s$ (terms of the form $\alpha_s^n L^{2n}$ and 
$\alpha_s^n L^{2n-1}$) can be calculated.  In \fig{eeFourJetFigure} we
also show the result of carrying out this resummation, and `matching'
the result to the fixed-order (NLO) calculation, which amounts to 
subtracting out common terms in the two expansions.   
This one-loop plus resummed result does improve the agreement with data 
still further (for $\mu=M_Z$) at small $\yc$.

Although it has been suppressed in \fig{eeFourJetFigure}, the NLO 
prediction for $R_4$ still has a reasonably large uncertainty stemming 
from the truncation of the perturbation series 
after just two terms, and the large size of the second term.  
(This uncertainty is perhaps 20-30\% if one estimates from the 
renormalization-scale dependence of the NLO result.) 
This feature limits the utility of $R_4$ in precision tests of QCD.
On the other hand, normalized distributions for various angles between the 
four jets tend to have quite small, and hence reliable, NLO 
corrections, of order 1-3\%.\cite{SignerMoriond,NTColorFactors}


\section{NNLO Predictions for Jet Physics?}

Uncertainties from higher-order terms still plague NLO calculations,
though not as severely as at Born level.
For example, $\alpha_s(M_Z^2)$ has been measured in $e^+e^-$ annihilation
at the $Z$ pole by using a dozen or more event-shape observables $O_i$ 
whose perturbative expansion begins at order $\alpha_s$ (such as
the three-jet fraction $R_3$):
$$
\eqalign{
O_i\ =\ A_i \left({\alpha_s(\mu)\over\pi}\right) &+ 
\left[ B_i + \coeff{1}{12} (33-2N_f) \ln(\mu^2/s) A_i \right] 
\left({\alpha_s(\mu)\over\pi}\right)^2 \cr
&+ \left[ ??? \right] \left({\alpha_s(\mu)\over\pi}\right)^3 + \cdots. \cr
}\equn\label{EventShapeExpansion}
$$
The extracted values of $\alpha_s(M_Z^2)$, obtained by fitting to the 
${\cal O}(\alpha_s^2)$ truncation of \eqn{EventShapeExpansion},
depend significantly on the value assumed for $\mu$ (an unphysical parameter),
and on the observable used, leading to a theory uncertainty that dominates the 
overall error,\cite{BurrowsReview}
$$
\alpha_s(M_Z^2)\ =\ 0.121 \pm 0.002(\hbox{exp.}) \pm 0.005(\hbox{theory}). 
\equn\label{AlphasEventShapes}
$$
Calculation of the next-to-next-to-leading order (NNLO) 
${\cal O}(\alpha_s^3)$ terms in \eqn{EventShapeExpansion} would clearly 
improve this situation.

To date, NNLO predictions in QCD are available for only a limited number of
rather inclusive quantities, such as the total cross-section in $e^+e^-$ 
annihilation.\cite{eeNNLO}  (Because such observables have a perturbative
expansion of the form $1 + \alpha_s/\pi + \cdots$, a competitive $\alpha_s$ 
extraction requires a very high precision measurement of the observable.) 
NNLO results for jet physics are still a ways off; however, some of the
ingredients are starting to be attacked.  For the above observables $O_i$,
three major classes of matrix elements are required:
\par\noindent
(1) tree-level $e^+e^-\to \gamma,Z \to$~5 partons,
\par\noindent
(2) one-loop $e^+e^-\to \gamma,Z \to$~4 partons,
\par\noindent
(3) two-loop $e^+e^-\to \gamma,Z \to$~3 partons.
\par\noindent
As we have seen, the first two classes are now available.  
There will also be a considerable amount of work required in order to 
reliably integrate the second, and particularly the first, class of
contributions over singular corners of phase space.  Some steps
have recently been taken along these lines.\cite{NigelUnresolved}

As for the two-loop matrix elements, an encouraging sign is that
a complete two-loop four-gluon scattering amplitude has been computed for 
the first time, using generalizations of the analytic tools described above, 
and the result is quite compact.\cite{BernRozowskyYan}
There are two caveats:  the theory is not QCD, but $N=4$ supersymmetric
$SU(N_c)$ Yang-Mills theory; and the answer has been expressed in terms of
scalar integrals, but those integrals have not yet been performed
in closed form.   The result for the leading-color part of the amplitude
is simply
$$
A_4^{N=4,\ \twoloop,\ {\rm L.C.}}\ =\ -N_c^2 \, st \, A_4^\tree 
  \left[ s \, \I_4(s,t) + t \, \I_4(t,s) \right],
\equn\label{TwoLoopN=4}
$$
where $\I_4(s,t)$ is the two-loop planar double-box scalar integral,
$$
\I_4(s,t) = \! \int \! {d^{D}p\ d^{D}q\over (2\pi)^{2D}}
 {1\over p^2  (p - k_1)^2 (p - k_1 - k_2)^2 (p + q)^2 q^2 
        (q-k_4)^2  (q - k_3 - k_4)^2 } \,.
\equn\label{I4Def}
$$
Whether two-loop results can be obtained in the same way for QCD remains
to be seen.


\section{Conclusions}

In charting the progress of particle
theory from the $S$-matrix days of the 1960s, to the triumph of the
Standard Model in the 1970s, to the recognition of its probable role
as an effective field theory in some grander theory,
Weinberg\,\cite{Weinberg} remarked:
\vskip0.2cm
\vbox{\leftskip=.2in\rightskip=.2in 
\noindent The justification of
any particular effective field theory is that it is simply the most
general possible theory that satisfies the axioms of analyticity,
unitarity, and cluster decomposition along with the relevant symmetry 
principles, so in a way our use today of effective field theories is the
ultimate revenge of $S$-matrix theory $\ldots$ Quantum field theory
is nothing but $S$-matrix theory made practical.
} 
\noindent
In some sense, the techniques reviewed here represent a further revenge of
$S$-matrix theory at the computational level.  Although the purely 
$S$-matrix-based bootstrap program of the 1960s proved intractable,
the lesson here is that a {\it perturbative} bootstrap program can
succeed.  Once helicity, color and supersymmetry decompositions are
used to break up loop amplitudes into components with simpler analytic
structure, this approach becomes a particularly efficient way to compute.   
As a practical result, improved QCD predictions for various
multi-jet processes are now emerging.



\section*{References}
\begin{thebibliography}{99}
\bibitem{NLOTwoJets} S.D.\ Ellis, Z. Kunszt and D.E.\ Soper,
         Phys.\ Rev.\ D40:2188 (1989);
         Phys.\ Rev.\ Lett.\ 64:2121 (1990);
         Phys.\ Rev.\ Lett.\ 69:1496 (1992)\semi
         F. Aversa, M. Greco, P. Chiappetta and J.P.\
         Guillet, Phys.\ Rev.\ Lett.\ 65:401 (1990)\semi
         F. Aversa, L. Gonzales, M. Greco, P. Chiappetta
         and J.P.\ Guillet, Z. Phys.\ C49:459 (1991)\semi
         W.T. Giele, E.W.N. Glover and D.A. Kosower,
         Phys.\ Rev.\ Lett. 73:2019 (1994);
         Phys.\ Lett.\ B339:181 (1994).
\bibitem{eeThreeJets} R.K. Ellis, D.A. Ross and A.E. Terrano,
         Phys.\ Rev.\ Lett. 45:1226 (1980);
         Nucl.\ Phys.\ B178:421 (1981)\semi
         K. Fabricius, I. Schmitt, G. Kramer and G. Schierholz,
         Phys.\ Lett.\ B97:431 (1980);
         Z.\ Phys.\ C11:315 (1981).
\bibitem{FiveGluon}  Z. Bern, L. Dixon and D.A. Kosower, Phys. Rev. Lett.
         70:2677 (1993).
\bibitem{Kunsztqqqqg} Z. Kunszt, A. Signer and Z. Tr\'ocs\'anyi, 
         Phys.\ Lett.\ B336:529 (1994).
\bibitem{qqggg} Z. Bern, L. Dixon and D.A. Kosower, 
         Nucl. Phys. B437:259 (1995).
\bibitem{TThreeJet} Z. Tr\'ocs\'anyi, Phys. Rev. Lett. 77:2182 (1996) 
         [hep-ph/9610499].
\bibitem{KGThreeJet} W.B. Kilgore and W.T. Giele, Phys. Rev. D55:7183
         (1997) [hep-ph/9610433]\semi
         W.B. Kilgore, in {\it Proceedings of the 1996
         A.P.S. D.P.F Meeting} [hep-ph/9609367]; 
         hep-ph/9705384, to appear in proceedings of 
         Moriond: QCD and High-Energy Hadronic Interactions, March 1997. 
\bibitem{Zqqqq} Z. Bern, L. Dixon, D.A. Kosower and S. Weinzierl, 
         Nucl. Phys. B489:3 (1997) [hep-ph/9610370].
\bibitem{ZqqggConf} Z. Bern, L. Dixon and D.A. Kosower, Nucl. Phys. Proc. 
         Suppl. 51C:243 (1996) [hep-ph/9606378].
\bibitem{Zqqgg} Z. Bern, L. Dixon and D.A. Kosower, hep-ph/9708239,
         to appear in Nucl.\ Phys.\ B.
\bibitem{GloverMiller} E.W.N. Glover and D.J. Miller, 
         Phys.\ Lett.\ B396:257 (1997) [hep-ph/9609474].
\bibitem{CGMqggq} J.M. Campbell, E.W.N. Glover and D.J. Miller, 
         Phys.Lett.B409:503 (1997) [hep-ph/9706297].
\bibitem{PrivateNigel} J.M. Campbell and E.W.N. Glover, 
         private communication.
\bibitem{FourJetsLC} A. Signer and L. Dixon, Phys. Rev. Lett. 78:811 (1997) 
         [hep-ph/9609460].
\bibitem{FourJets} A. Signer and L. Dixon, Phys. Rev. D56:4031 (1997) 
         [hep-ph/9706285].
\bibitem{NTFourJets} Z. Nagy and Z. Tr\'ocs\'anyi, Phys. Rev. Lett. 79:3604
         (1997) [hep-ph/9707309]; hep-ph/9708344, to appear in proceedings
         of QCD 97, July 1997.
\bibitem{SignerMoriond} A. Signer, hep-ph/9705218, to appear in 
         proceedings of Moriond: QCD and High-Energy Hadronic 
         Interactions, March 1997.
\bibitem{NTColorFactors} Z. Nagy and Z. Tr\'ocs\'anyi, hep-ph/9712385.
\bibitem{AnnualReview} Z. Bern, L. Dixon and D.A. Kosower,
         Ann. Rev. Nucl. Part. Sci. 46:109 (1996) [hep-ph/9602280]. 
\bibitem{TASIReview} L. Dixon, in 
         {\it QCD \& Beyond: Proceedings of TASI '95}, 
         ed. D.E.\ Soper (World Scientific, 1996) [hep-ph/9601359].
\bibitem{OtherReviews} Z. Bern, L. Dixon and D.A. Kosower, 
         Nucl. Phys. Proc. Suppl. 51C:243 (1996) [hep-ph/9606378];
         L. Dixon, hep-ph/9507214, in {\it Proceedings of SUSY95}, 
         eds I. Antoniadis and H. Videau (Editions Frontieres, 1996);
         Z. Bern, L. Dixon, D.C. Dunbar and D.A. Kosower,
         hep-ph/9706447, to appear in proceedings of DIS 97.
\bibitem{ManganoReview}
         M. Mangano and S.J. Parke, Phys.\ Rep.\ 200:301 (1991).
\bibitem{Long}
         Z. Bern and D.A.\ Kosower, Nucl.\ Phys.\ 379:451 (1992).
\bibitem{KunsztFourPoint} Z. Kunszt, A. Signer and Z. Tr\'ocs\'anyi, 
         Nucl.\ Phys.\ B411:397 (1994). 
\bibitem{Bootstrap} S. Mandelstam, Phys. Rev. 112:1344 (1958); 
         G.F. Chew and S.C. Frautschi, Phys. Rev. 123:1486 (1961);
         G.F. Chew and A. Pignotti, Phys. Rev. 176:2112 (1968). 
\bibitem{RecursiveBG}
         F.A.\ Berends and W.T.\ Giele, Nucl.\ Phys.\ B306:759 (1988).
\bibitem{Recursive}
         D.A.\ Kosower, Nucl.\ Phys.\ B335:23 (1990);
         G. Mahlon and T.-M.\ Yan,  Phys.\ Rev.\ D47:1776 (1993), 
         hep-ph/9210213; 
         G. Mahlon, Phys.\ Rev.\ D47:1812 (1993), hep-ph/9210214;
         G. Mahlon, T.-M.\ Yan and C. Dunn, Phys.\ Rev.\ D48:1337 (1993)
         [hep-ph/9210212].
\bibitem{SusyFour} Z. Bern, L. Dixon, D.C. Dunbar and D.A. Kosower,
         Nucl.\ Phys.\ B425:217 (1994), hep-ph/9403226.
\bibitem{BCFactorization}
         Z. Bern and G. Chalmers, Nucl.\ Phys.\ B447:465 (1995), 
         hep-ph/9503236;
         G. Chalmers, hep-ph/9405393, in {\it Proceedings of the XXII 
         ITEP International Winter School of Physics} 
         (Gordon and Breach, 1995).
\bibitem{SpinorHelicity}
         P.\ De Causmaecker, R.\ Gastmans, W.\ Troost and T.T.\ Wu,
         Phys.\ Lett.\ 105B:215 (1981), Nucl.\ Phys.\ B206:53 (1982)\semi
         R.\ Kleiss and W.J.\ Stirling, Nucl.\ Phys.\ B262:235 (1985)\semi
         R.\ Gastmans and T.T.\ Wu,
         {\it The Ubiquitous Photon: Helicity Method for QED and QCD}
         (Clarendon Press, 1990)\semi
         Z. Xu, D.-H.\ Zhang and L. Chang, Nucl.\ Phys.\ B291:392 (1987).
\bibitem{OldSWI}
         M.T.\ Grisaru, H.N.\ Pendleton and P.\ van Nieuwenhuizen,
         Phys. Rev. {D15}:996 (1977);
         M.T. Grisaru and H.N. Pendleton, Nucl.\ Phys.\ B124:81 (1977).
\bibitem{NewSWI}
         S.J. Parke and T. Taylor, Phys.\ Lett.\ 157B:81 (1985);
         Z. Kunszt, Nucl.\ Phys.\ B271:333 (1986). 
\bibitem{SusyOne}
         Z. Bern, L. Dixon, D.C. Dunbar and D.A. Kosower,
         Nucl.\ Phys.\ B435:39 (1995) [hep-ph/9409265].
%
\bibitem{TreeSixGluon} S. Parke and T. Taylor, 
         Nucl. Phys. B269:410 (1986);
         Z. Kunszt, Nucl. Phys. B271:333 (1986);
         J. Gunion and J. Kalinowski, Phys. Rev. D34:2119 (1986);
         F.A. Berends and W.T. Giele, Nucl. Phys. B294:700 (1987);
         M. Mangano, S. Parke and Z. Xu, Nucl. Phys. B298:653 (1988).
\bibitem{ppkt} S. Catani, Yu.L. Dokshitser and B.R. Webber, Phys. Lett. 
         B285:291 (1992); S. Catani, Yu.L. Dokshitser, M.H. Seymour and 
         B.R. Webber, Nucl. Phys. B406:187 (1993); 
         see also S.D. Ellis and D. Soper, Phys. Rev. D48:3160 (1993) 
         [hep-ph/9305266].
\bibitem{FixedCone} UA2 Collab., Phys. Lett. B257:232 (1991).
\bibitem{IterativeCone} CDF Collab., Phys. Rev. D45:1448 (1992);
         D0 Collab., Phys. Rev. D53:6000 (1996) [hep-ex/9509005].
\bibitem{EKSalg} S.D. Ellis, Z. Kunszt and D.E. Soper, 
         Phys. Rev. Lett. 62:726 (1989).
\bibitem{SeymourIR} M.H. Seymour, hep-ph/9707338.
%
\bibitem{MENLOPARC} A. Signer, Comput.\ Phys.\ Commun.\ 106:125 (1997). 
\bibitem{BerGie89} 
         F.A. Berends, W.T. Giele and H. Kuijf, Nucl.\ Phys.\ B321:39 (1989). 
\bibitem{FiveJetsBorn} 
         K. Hagiwara and D. Zeppenfeld, Nucl.\ Phys.\ B313:560 (1989); 
         N.K. Falck, D. Graudenz and G. Kramer, Nucl. Phys. B328:317
         (1989). 
\bibitem{ERT} R.K. Ellis, D.A. Ross and A.E. Terrano,
         Phys.\ Rev.\ Lett. 45:1226 (1980);
         Nucl.\ Phys.\ B178:421 (1981).
\bibitem{KunsztSoperFKSCS} Z. Kunszt and D.E. Soper, Phys.\ Rev.\ D46:192 
        (1992);
        S. Frixione, Z. Kunszt and A. Signer, Nucl.\ Phys.\ B467:399
        (1996) [hep-ph/9512328];
        S. Catani and M.H. Seymour, Phys.\ Lett.\ B378:287 (1996) 
        [hep-ph/9602277], Nucl.\ Phys.\ B485:291 (1997) [hep-ph/9605323].
\bibitem{Durham} N. Brown and W.J. Stirling, Z. Phys. C53:629 
         (1992).
\bibitem{DurhamResum} S. Catani, Yu.L. Dokshitser, M. Olsson, G. Turnock 
         and B.R. Webber, Phys. Lett. B269:432 (1991).
\bibitem{ALEPHReview} ALEPH Collab., R. Barate et al., preprint 
         CERN-PPE-96-186, to appear in Phys. Rep.
\bibitem{SLDdata} SLD Collab., K. Abe et al., Phys. Rev. Lett. 
         71:2528 (1993); P. Burrows (SLD Collab.), 
         private communication.
\bibitem{BurrowsReview} For a review, see P. Burrows, Acta Phys.\ Polon.\ 
         B28:701 (1997) [hep-ex/9612007].
\bibitem{eeNNLO}
         S.G.\ Gorishny, A.L.\ Kataev and S.A.\ Larin, Phys.\ Lett.\ 
         B259:144 (1991); 
         L.R. Surguladze and M.A.\ Samuel, Phys.\ Rev.\ Lett. 66:560 (1991); 
         erratum {\it ibid.} 66:2416 (1991).    
\bibitem{NigelUnresolved} 
         A. Gehrmann-De Ridder and E.W.N. Glover, hep-ph/9707224; 
         J.M. Campbell and E.W.N. Glover, hep-ph/9710255.
\bibitem{BernRozowskyYan} Z. Bern, J.S. Rozowsky and B. Yan, 
         Phys.\ Lett.\ B401:273 (1997) [hep-ph/9702424]; 
         hep-ph/9706392, to appear in proceedings of DIS 97.
\bibitem{Weinberg} S. Weinberg, in {\it The Rise of the Standard Model}, 
         eds. L. Hoddeson, L. Brown, M. Riordan and M. Dresden 
         (Cambridge, 1997).
\end{thebibliography}
\end{document}